# Impact of plasma potential and plasma sheath composition on the InN PA-MOCVD growth kinetics and structural properties

*Zaheer Ahmad, Alexander Kozhanov*

In this paper, we present the effect of various growth parameters of PA-MOCVD on the plasma potential. The impact of varying plasma potential on the growth and structural properties of the InN films, and how the growth process and film properties will be affected if an external DC bias perturbs the plasma potential.

**Plasma potential and the plasma sheath**

Any gas-phase plasma consists of electrons, ions, and neutrals. In a plasma, the densities of electrons and ions are the same on average so that the plasma as a whole is electrically neutral. In RF plasma, owing to their smaller mass, the electron speed is much larger[1] than the ions on average. The number of electrons and ions is the same in the plasma, which, by definition, is electrically neutral. Hence, the charge distribution between the plasma source and the grounded substrate is not uniform: more electrons reach substrate per unit time than ions.

Consequently, a negative potential builds upon the substrate with respect to the plasma. This potential at the substrate affects the motion of ions and electrons near it. Therefore, in normal circumstances, the bulk of the plasma is positively charged with respect to the substrate. The plasma sheath, thus, represents the non-neutral potential region between the plasma and the substrate. Ions and electrons recombine at the substrate and leave the plasma system.

The difference in potential of the bulk plasma and the substrate gives rise to an electric field that accelerates ions towards the substrate. A grounded substrate will draw currents from the plasma; otherwise, the substrate floats at a potential. The plasma is essentially quasi-neutral and is at a constant positive potential ($V_P$). The floating potential developed at the isolated substrate can be termed as $V_f$. If the substrate is an insulator, different parts of its surface may be at different floating potentials due to the difference in the distribution of electron speeds in the local area surrounding the plasma and the electric field set up in the insulator itself.

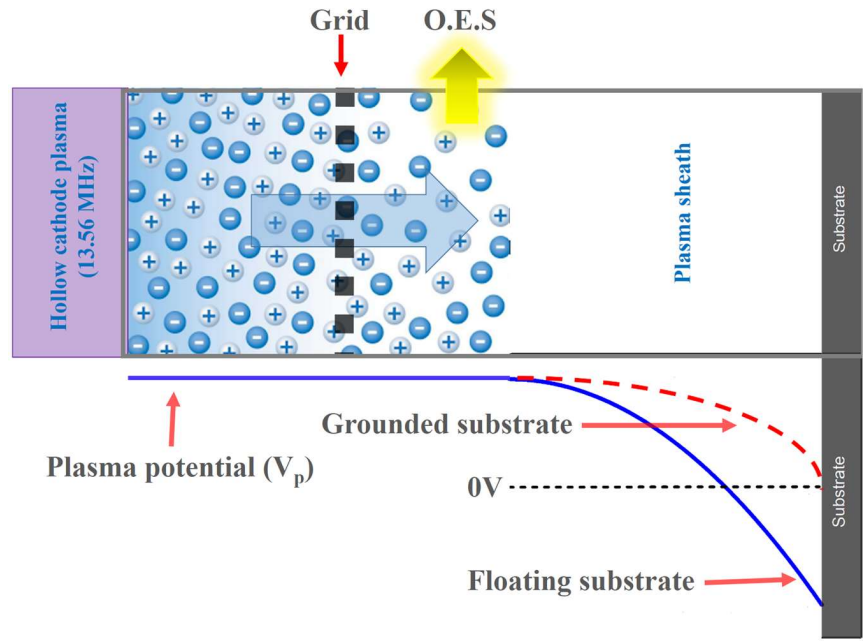

**Figure 1**: Schematic of the electric potential distribution in the remote PA-MOCVD system. OES represents the optical emission spectroscopy (distances are not to the scale)

**Plasma sheath dynamics**

A thorough understanding of the dynamics of plasma sheath formed over a substrate by an RF plasma is essential in the fundamental as well as practical aspects. The plasma sheath dynamics control the energy and distribution of ions reaching the substrate, which consequently affects the surface reaction rate, surface morphology, crystalline, and electrical properties of the resulting films grown. High-density plasma reactors have the advantage of quasi-independent control of plasma density and ion bombardment energy[2]. In PECVD systems, it can be achieved by separating plasma production (Hollow cathode plasma source operating at 13.56 MHz in our case) from the bias voltage applied to the substrate or a grid in between the plasma source and the substrate. The effect of grid bias on the PA-MOCVD process will be discussed later in the results section of this chapter.

A plasma sheath model developed by Theodoros and Demetre for a collisionless sheath[3] has been used in the present analysis. The variation in potential within the sheath is described by the Poisson's equation:

$$\nabla^2 V = -\frac{\rho}{\epsilon_o}, \quad (1)$$

where V is potential, $\rho$ is the charge density in plasma, and $\epsilon_o$ is the permittivity of free space. The density $\rho$ in the plasma may be expressed as

$$\rho = e(n_i - n_e), \quad (2)$$

where $e$ is the electron charge, $n_i$ is the ion density, and $n_e$ is the electron density. Following the Boltzmann distribution, the electron density is given by:

$$n_e(x,t) = n_e(x_s,t) \exp\left(\frac{V(x,t)-V(x_s,t)}{k_B T_e}\right), \quad (3)$$

where $x_s$ represents the sheath edge, $T_e$ is the electron temperature. The determination of $n_e$ and $T_e$ from emission spectra is based upon the assumption that electrons have a Maxwellian energy distribution. The electron temperature was also assumed to be constant in the bulk of the plasma, both spatially and temporally. Moreover, if the sheath is considered to be collisionless, the ion density $n_i$ and ion fluid velocity $u_i$ can be described by the following conservation equations.

$$n_i(x,t)u_i(x,t) = n_i(x_s,t)u_i(x_s,t) \quad (4)$$

$$\frac{m_i}{2}u_i^2(x,t) + e\bar{V}(x,t) = \frac{m_i}{2}u_i^2(x_s,t) + e\bar{V}(x_s,t) \quad (5)$$

Equation 4 is based upon the assumption of adiabatic ions[4]. Equations 4 and 5 can be solved for the ion density:

$$n_i(x,t) = n_i(x_s,t)(1 - \frac{2e}{m_i u_i^2(x_s,t)}(\bar{V}(x,t) - \bar{V}(x_s,t)))^{-\frac{1}{2}} \quad (6)$$

Ions respond to an effective damped potential $\bar{V}$ (or field) rather than the actual potential $V$.

$$\frac{\partial \bar{V}}{\partial t}(x,t) = \frac{\bar{V}(x,t) - V(x,t)}{\tau_i} \quad (7)$$

where $\tau_i$ is the time that ions take to transit through the sheath. In Equation 7 $\tau_i$ acts as a time constant that controls the time averaging of $\bar{V}$. Numerically $\tau_i$ is the inverse of the ion plasma frequency.

$$\tau_i = \frac{1}{\omega_{Pi}} = \sqrt{\frac{\epsilon_o m_i}{e^2 n_s}} \quad (8)$$

Here, $n_s$ is the plasma density at the sheath edge.

Equations 2, 3, 6, and 7 may now be substituted back into Equation 1 to get the following equations that describe the sheath dynamics.

$$-\frac{e(n_i(x) - n_e(x))}{\epsilon_o} =$$

$$\frac{en_e(x_s)}{\epsilon_o}\left\{\exp\left(\frac{V(x,t) - V(x_s,t)}{k_B T_e}\right) - \left\{1 - \frac{2e}{m_i u_i^2(x_s,t)}(\bar{V}(x,t) - \bar{V}(x_s,t))\right\}^{-\frac{1}{2}}\right\} \quad (9)$$

Equation 9 governs the sheath potential and ion and electron densities.

The sheath would be a positive space charge region only if

$$\nabla^2 V < 0; \quad x > 0 \quad (10)$$

and

$$\nabla^2 V = 0; \quad x = 0 \quad (11)$$

Condition 10 implies that

$$\left\{1 - \frac{2e}{m_i u_i^2(x_s,t)}\left(\bar{V}(x,t) - \bar{V}(x_s,t)\right)\right\}^{-\frac{1}{2}} > \exp\left(\frac{V(x,t)-V(x_s,t)}{k_B T_e}\right),$$

which can be simplified to

$$u_i(x_s, t) > \sqrt{\frac{k_B T_e}{m_i}} \quad (12)$$

Condition 10 implies that only the ions that have a velocity $\sqrt{\frac{k_B T_e}{m_i}}$ or greater can enter the sheath. This velocity is called Bohm velocity, and condition 12 is the Bohm criterion.

$$u_B = \sqrt{\frac{k_B T_e}{m_i}} \quad (13)$$

The Bohm criterion guarantees the ion flux towards any object that is negatively charged with respect to the plasma. The ions thus have to enter the sheath with a velocity that is much larger than the thermal velocity. Hence there must be an electric field in the quasi-neutral region of the plasma that accelerates ions to an energy of at least $\frac{1}{2}k_B T_e$ towards the sheath edge. This field is usually associated with the collisions, ionization, or other sources of particle collisions. This region is also referred to as the presheath, which extends over distances of the order of plasma dimensions, and the presheath field is weak enough not to violate the quasi-neutrality condition.

**Sample growth, plasma potential measurement, and emission spectra acquisition**

Four InN film sample sets (labeled as A, B, C, and D) were grown using PA-MOCVD on c-plane $Al_2O_3$ wafers offcut at 0.2° towards m-plane. Trimethylindium (TMI) and nitrogen plasma were used as group-III and group-V precursors, respectively. The nitrogen plasma was produced using an RF hollow cathode plasma source. Nitrogen gas was used as the TMI carrier gas.

The substrate preparation steps for four sets were kept same and are as follows. Sapphire substrates were heated to 250°C. At 250°C the substrates were exposed to 100sccm of $H_2$ gas at 2.1Torr for 5 minutes. After that, the substrates were further heated to 500°C in nitrogen gas keeping the pressure at 2.1Torr. The InN nucleation layer was deposited at 500°C, 2.1Torr, 150W RF power, 750sccm $N_2$ plasma and 250sccm Trimethylindium for 5 mins. The samples were further heated to 775°C. The unconventional high-temperature growth of InN has been previously demonstrated by our group[5]. The growth conditions used for each of the sample sets are summarized in Tables 1, 2, 3, and 4.

| TMI µ-mol/min | Plasma-$N_2$ µ-mol/min | Growth pressure(Torr) | RF power (W) | Grid bias (V) | Growth Temperature(°C) |
|---|---|---|---|---|---|
| 1.03 | 484.5 | **2.8** | 400 | Float | 775 |
| 1.03 | 484.5 | **3.4** | 400 | Float | 775 |
| 1.03 | 484.5 | **3.8** | 400 | Float | 775 |
| 1.03 | 484.5 | **4.1** | 400 | Float | 775 |
| 1.03 | 484.5 | **4.8** | 400 | Float | 775 |

**Table 2.** Growth conditions for set B; the pressure series.

| TMI µ-mol/min | Plasma-$N_2$ µ-mol/min | Growth pressure(Torr) | RF power (W) | Grid bias (V) | Growth Temperature(°C) |
|---|---|---|---|---|---|
| 8.39 | 484.5 | 2.2 | **100** | Float | 775 |
| 8.39 | 484.5 | 2.2 | **150** | Float | 775 |
| 8.39 | 484.5 | 2.2 | **200** | Float | 775 |
| 8.39 | 484.5 | 2.2 | **250** | Float | 775 |
| 8.39 | 484.5 | 2.2 | **300** | Float | 775 |

**Table 3.** Growth conditions for set C; the RF power series.

Optical emission spectroscopy was employed to measure a range of the plasma parameters, including the fluxes and densities of neutral/ionic species and the plasma temperature[6,7,8]. The emission spectra were measured as close as possible to the growth surface via a quartz window. The floating potential at the grid is the result of a continuous bombardment of the various plasma species onto the g rid. It is the function of different process parameters such as chamber pressure, nitrogen flow through plasma source, and the plasma power. The floating potential was externally measured at the grid for each set of samples.

| TMI µ-mol/min | Plasma-$N_2$ µ-mol/min | Growth pressure(Torr) | RF power (W) | Grid bias (V) | Growth Temperature(°C) |
|---|---|---|---|---|---|
| 4.25 | 37.25 | 0.35 | 50 | **-100** | 775 |
| 4.25 | 37.25 | 0.35 | 50 | **-50** | 775 |
| 4.25 | 37.25 | 0.35 | 50 | **ground** | 775 |
| 4.25 | 37.25 | 0.35 | 50 | **+50** | 775 |
| 4.25 | 37.25 | 0.35 | 50 | **+100** | 775 |

**Table 4.** Growth conditions for set D; the grid bias series.

For the sample set A (4.12Torr, 400W, 1.03µ-mol/min trimethyl indium), the input Plasma $N_2$ flow was varied in the range 447 – 931 µ-mol/min. For the set B (400W, 484 µ-mol/min Plasma $N_2$, 1.03µ-mol/min trimethyl indium), the reactor pressure was varied in the range 2.8 – 4.8 Torr. For the set C (2.2Torr, 484 µ-mol/min Plasma $N_2$, 8.39µ-mol/min trimethyl indium), the RF power vas varied in the range 100 – 300 W. For the set D of samples, all the growth parameters were kept constant (0.35Torr, 50W, 37.25µ-mol/min Plasma $N_2$, 4.25µ-mol/min trimethyl indium) and the grid was externally biased with a DC potential. The grid bias was varied from -100V to +100V.

## Correlation of plasma potential ($V_P$) with various plasma species

All the plasma emission spectra were analyzed using the methods of atomic and molecular spectroscopy, using the Boltzmann equation and Saha equation following the method described elsewhere[8]. The results of the plasma analysis for each of the grown samples are summarized in Fig. 2. The variation in the density of atomic nitrogen ions is found to be in good correlation with the change in the plasma potential measured at the grid. However, no correlation is found between the plasma potential and the variation in atomic neutrals or molecular plasma species.

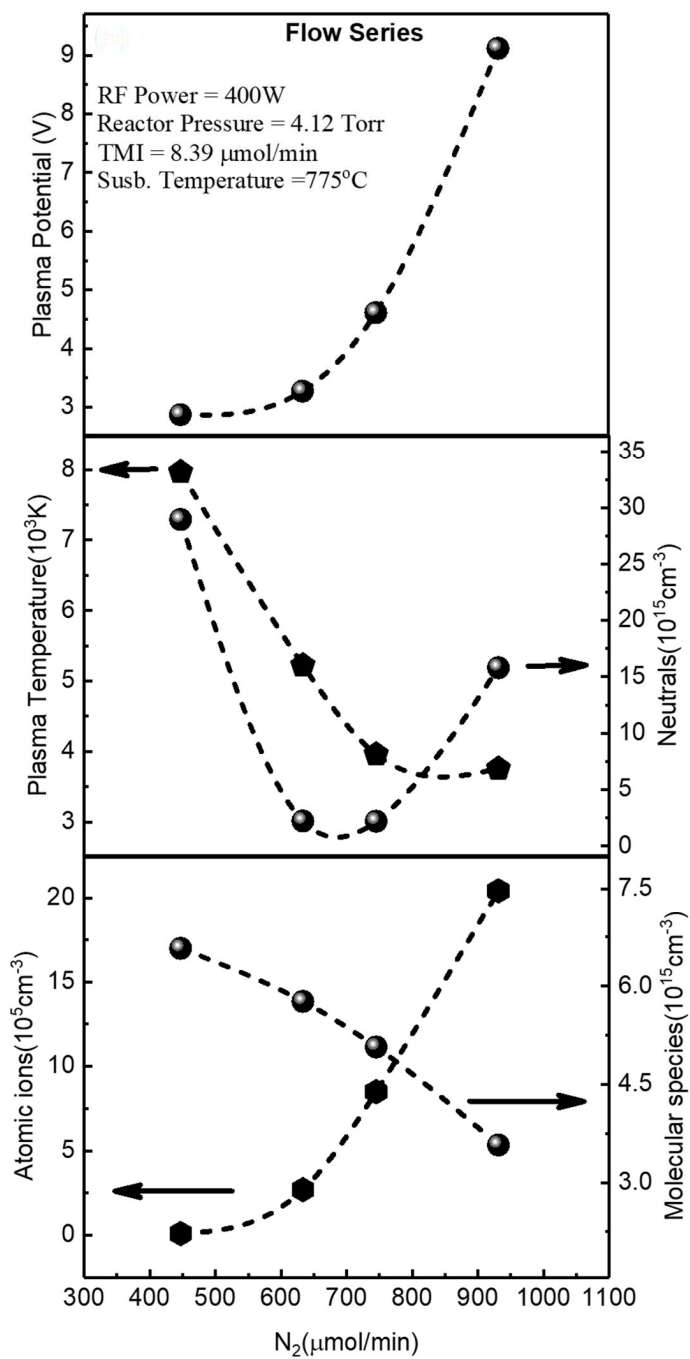

**Figure 2**: Variation in plasma potential and plasma composition with plasma $N_2$ flow rate.

For the sample set A (4.12Torr, 400W, 1.03μ-mol/min TMI), the molecular nitrogen species density, consisting of both metastable and ionized molecular nitrogen, decreases with increasing input nitrogen flow through the plasma source. The plasma temperature (average ion energy) also decreases with an increase in the input nitrogen flow as the same energy given by the plasma source is now distributed among more of the species than before. The atomic nitrogen ions density increases with input nitrogen at the given conditions.

For the sample set B (400W, 484 μ-mol/min Plasma $N_2$, 1.03μ-mol/min TMI), the molecular nitrogen species density increases upon increasing the chamber pressure. With increasing pressure, the collisional quenching and diffusion induced recombinations become the 8$N_2$ molecules[9]. These two processes cause a decrease in atomic nitrogen ions.

For the set C (2.2Torr, 484 μ-mol/min Plasma $N_2$, 8.39μ-mol/min TMI), the molecular nitrogen species density, consisting of both metastable and ionized molecular nitrogen, increases with increasing RF power up to 200 W as more nitrogen can be excited by electrons. However, at high RF power, molecular nitrogen species density decreases due to the collisional quenching with other $N_2$ molecules. The collisional quenching becomes the dominant de-excitation process at higher pressures[9]. In contrast, both neutral and ionized atomic nitrogen loss through N–N gas-phase recombination require three-body collisions, which do not begin to compete with the loss by diffusion to the chamber walls until higher pressures[9]. Since diffusion losses decrease with increasing pressure, ionized atomic nitrogen density decreases with the chamber pressure at constant RF power. It can be concluded from the emission spectra and plasma potential recorded at the grid for sets A, B and C that the plasma potential is mainly correletaed with the concentration of atomic nitrogen ions. The dependence of $V_P$ on the density of atomic nitrogen ions is shown in Fig. 6.

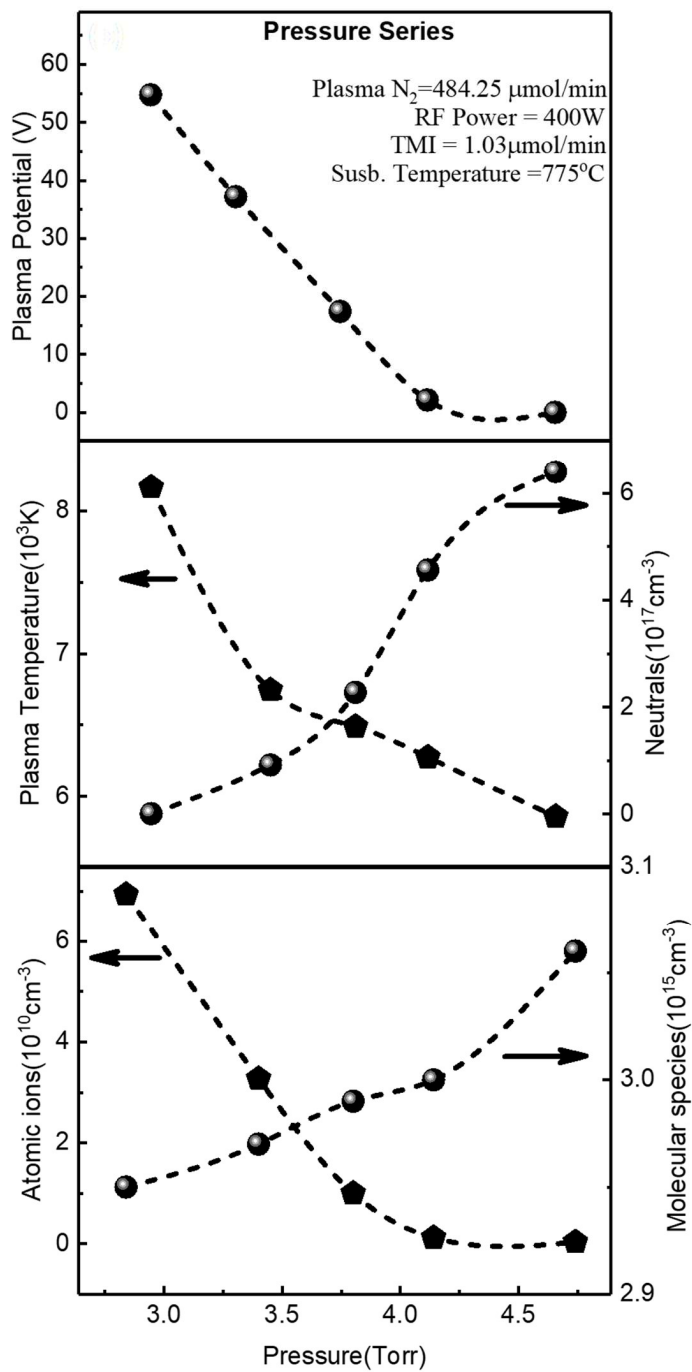

**Figure 3**: Variation in plasma potential and plasma composition with reactor pressure.

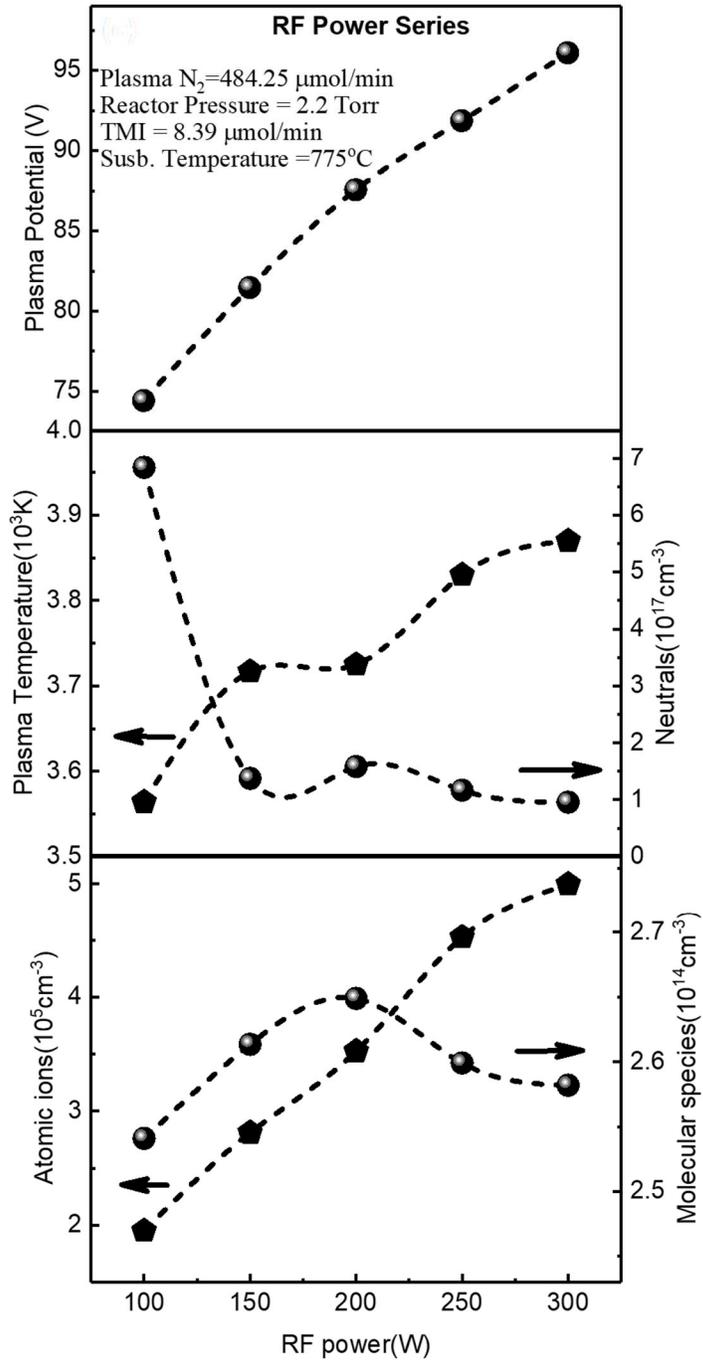

**Figure 4**: Variation in plasma potential and plasma composition with plasma RF power.

For the sample set D (0.35Torr, 50W, 37.25μ-mol/min Plasma $N_2$, 4.25μ-mol/min trimethyl indium), the density molecular species (consisting of both metastable and ionized molecular nitrogen) and of atomic nitrogen ions decreases with increasing grid bias. The plasma temperature (average ion energy) also decreases with an increase in the grid bias. This is due to the fact that positive ions are bneing repelled by the increased potential on the grid.

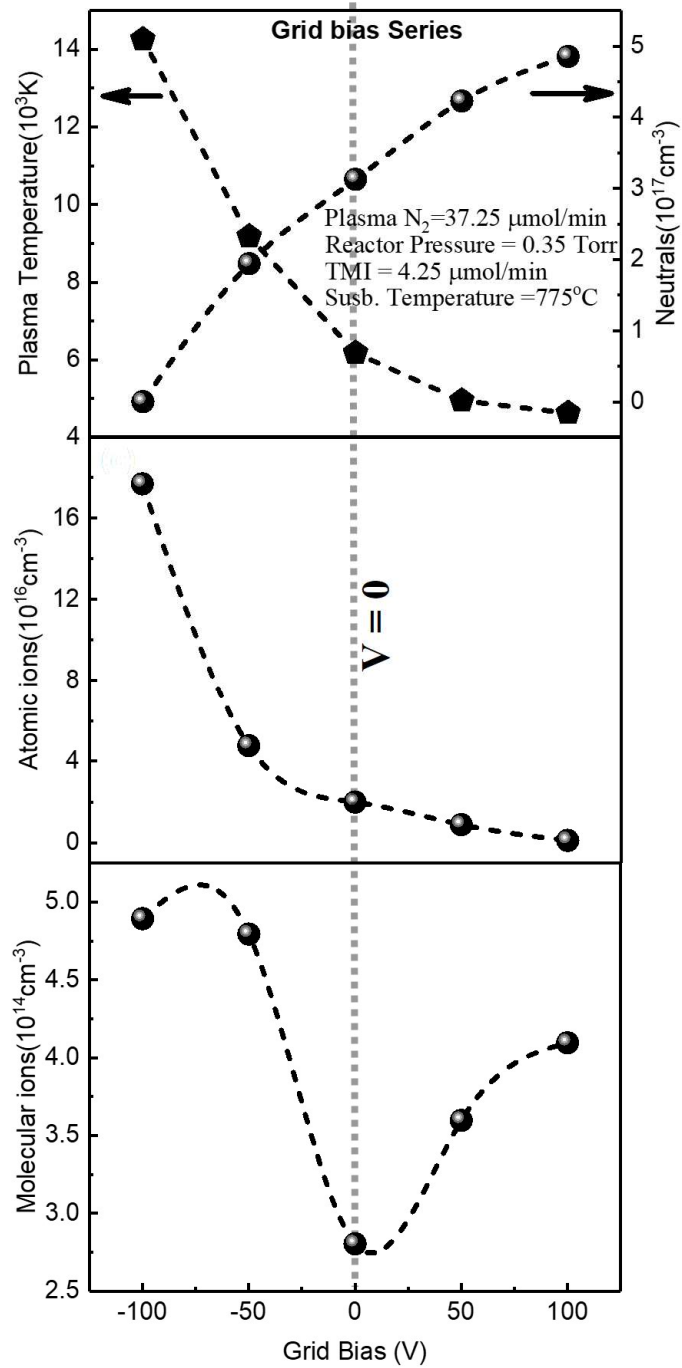

**Figure 5**: Variation in plasma composition with plasma grid bias. V = 0 at the dotted line.

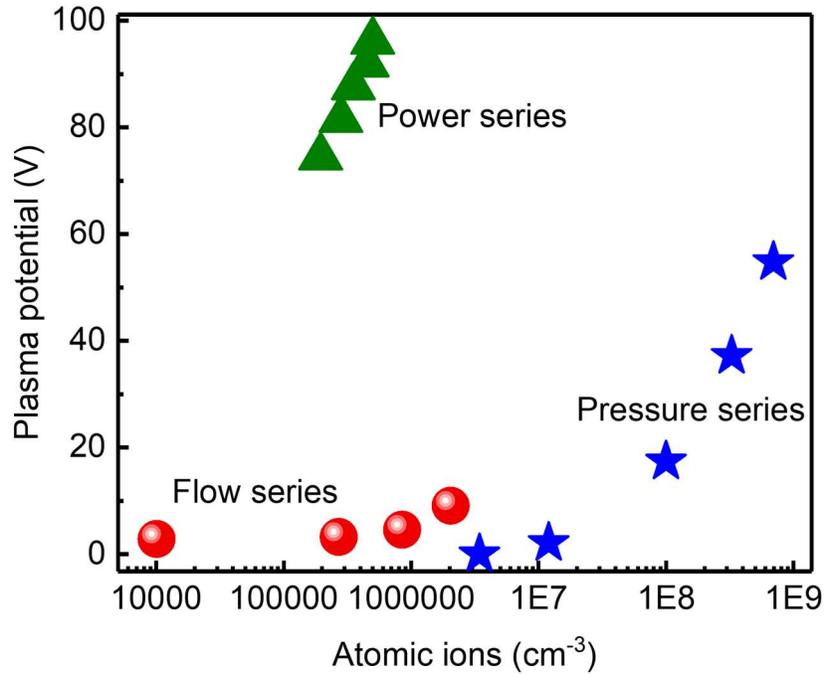

**Figure 6**: The dependence of plasma potential on the concentration of atomic ions.

**The impact of plasma potential on structural properties of InN**

The influence of plasma potential and grid bias on the structural properties of InN was studied via Raman spectroscopy. Raman spectra for the four sets of InN samples grown at varying conditions are shown in Fig. 3. The $E_2$-high (489 cm$^{-1}$) and $A_1$-LO (590 cm$^{-1}$) phonon modes were observed in each of the sample sets.

The phonon relaxation time was estimated using the FWHM and peak position of the $E_2$-high and A1-LO phonon modes extracted from a multi-peak fit of the Raman spectra. Fig. 4 shows the $E_2$-high phonon mode relaxation time as a function of plasma potential for each set of samples. The $E_2$-high phonon relaxation time increases with the increase in plasma potential within each of the sample sets. Since the plasma potential is directly related to the concentration of atomic nitrogen ions and these species are the primarily contributing[8] to the InN growth, the increase in $E_2$-high phonon relaxation time is likely to be caused by a decrease in nitrogen vacancies in the

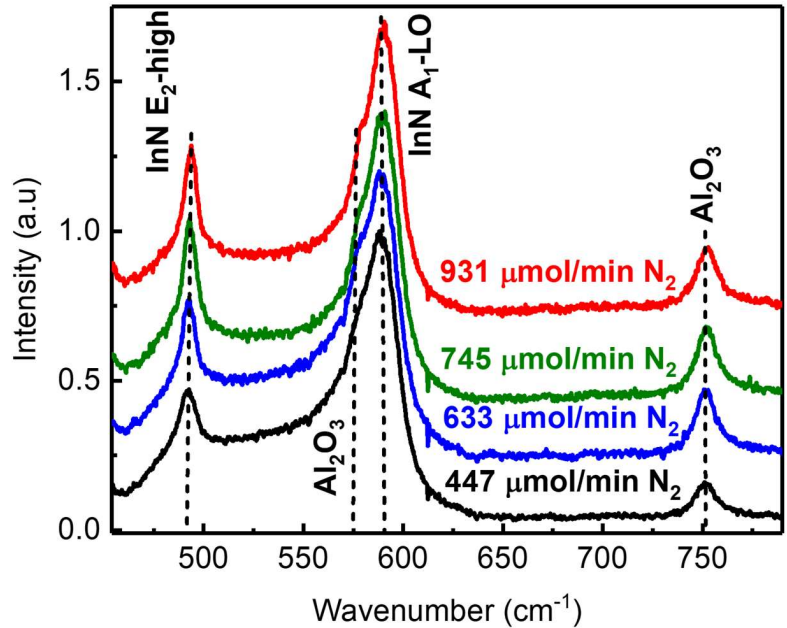

**Figure 7**: Raman spectra for the sample set A grown InN, which are primary contributors to the $E_2$-high mode[10]. No significant correlation has been observed between the $A_1$-LO phonon mode relaxation time and the plasma potential

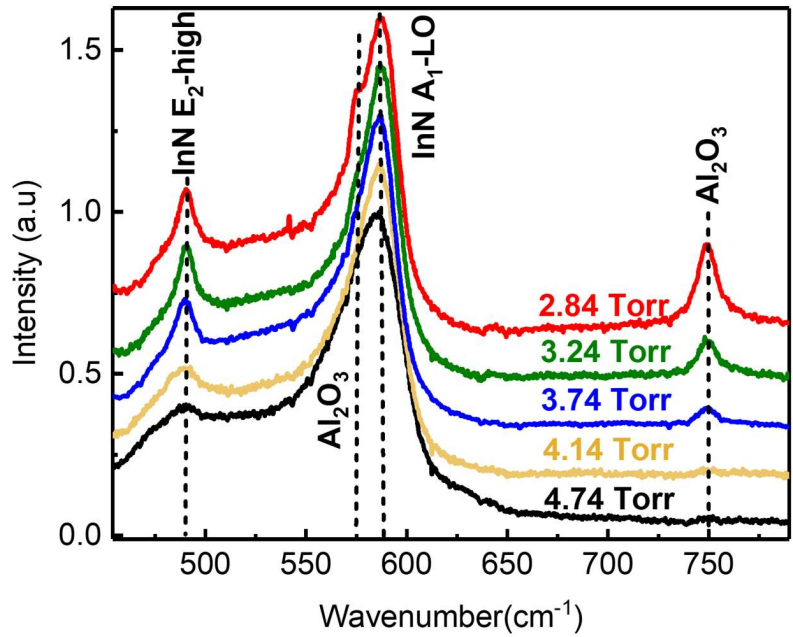

**Figure 8**: Raman spectra for the sample set B

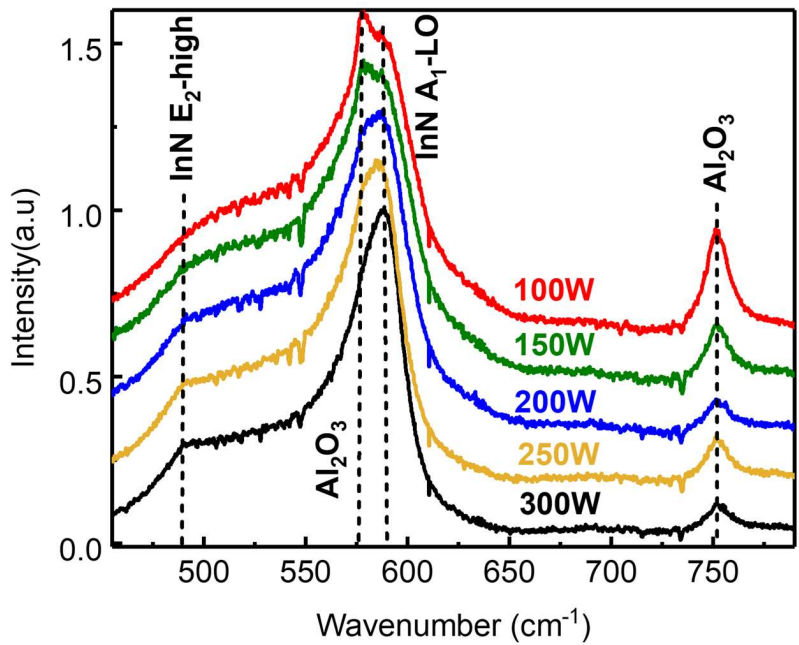

**Figure 9**: Raman spectra for the sample set C

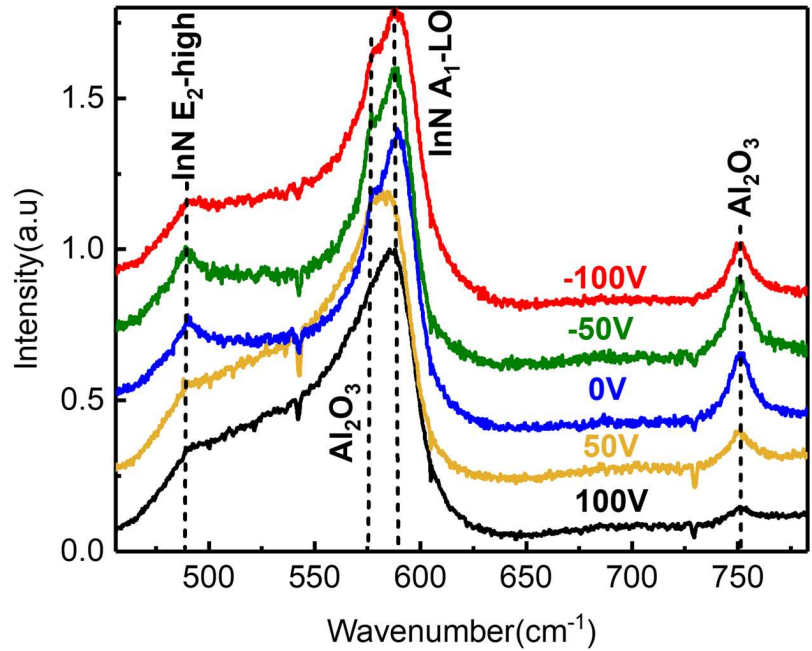

**Figure 10**: Raman spectra for the sample set D

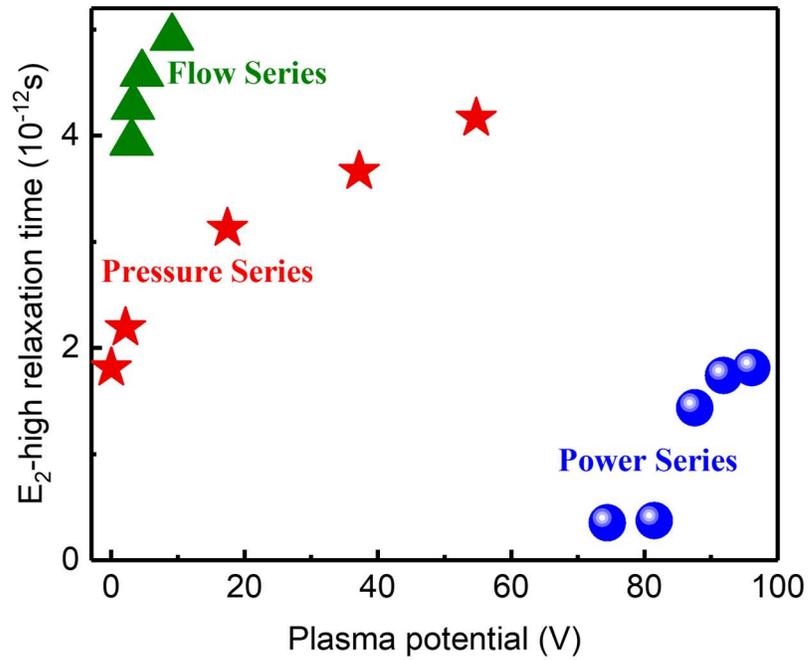

**Figure 11**: The impact of plasma potential on the $E_2$-high phonon mode relaxation time.

**Conclusion**

The plasma potential is found to be in a positive correlation with the concentration of atomic nitrogen ions. For any set of growth parameters, the increase in plasma potential is due to an increase in the concentration of atomic nitrogen ions, and that, in turn, causes to improve the crystalline quality of the InN films grown. These studies also establish the fact that the plasma potential can be used as a qualitative measure of the concentration of atomic nitrogen ions.